# Monte Carlo simulation of prompt γ-ray spectra from depleted uranium under D−T neutron irradiation and electron recoil spectra in a liquid scintillator detector *

QIN Jian-Guo(秦建国)[1,2,3] LAI Cai-Feng(赖财锋)[2]  LIU Rong(刘荣)[2]

ZHU Tong-Hua(朱通华)[2;1)]   ZHANG Xin-Wei(张信威)[4]   YE Bang-Jiao(叶邦角)[1]

[1] Department of Modern Physics, University of Science and Technology of China, Hefei 230026, China

[2] Institute of Nuclear Physics and Chemistry, China Academy of Engineering Physics, P.O.Box 213, Mianyang 621900, China

[3] Graduate School, China Academy of Engineering Physics, Mianyang 621900, China

[4] Institute of Applied Physics and Computational Mathematics, P.O.Box 8009, Beijing 100088, China

**Abstract:** To overcome the problem of inefficient computing time and unreliable results in MCNP5 calculation, a two-step method is adopted to calculate the energy deposition of prompt γ-rays in detectors for depleted uranium spherical shells under D–T neutrons irradiation. In the first step, the γ-ray spectrum for energy below 7 MeV is calculated by MCNP5 code; secondly, the electron recoil spectrum in a BC501A liquid scintillator detector is simulated based on EGSnrc Monte Carlo Code with the γ-ray spectrum from the first step as input. The comparison of calculated results with experimental ones shows that the simulations agree well with experiment in the energy region 0.4–3 MeV for the prompt γ-ray spectrum and below 4 MeVee for the electron recoil spectrum. The reliability of the two-step method in this work is validated.

## 1 Introduction

Prompt radiation measurement is an essential method in the diagnostics of nuclear explosions [1]. In addition, prompt γ-ray spectra can give important support in validating nuclear cross sections [2, 3]. In the physical design of hybrid reactors, the radiation effects and $^{238}$U nuclear parameters have to be taken into account to guarantee the feasibility of the design. Benchmark data from experiments on macroscopic samples or mock-ups of depleted uranium can be used to evaluate cross sections and check the related nuclear data. The representative experiment is the pulsed spheres program at LLNL [4–7], in which the electron recoil spectrum (ERS) in a liquid scintillator detector was investigated for prompt γ-rays from a depleted uranium spherical sample under D–T neutron irradiation.

Monte Carlo (MC) simulation and comparison with experiment is a necessary validation method. To simulate the energy deposition of prompt γ-rays induced by D–T neutrons, the coupled neutron-photon Monte Carlo transport mode is needed. MCNP5 is one of the most general codes that can do this work with a pulse height tally. However, the pulse height tally does not work well with neutrons, and a very long computing time necessary to simulate the whole macroscopic experimental model. Furthermore, the calculation is not reliable as the intrinsic non-analogous nature of neutron transport in MCNP [8], as was encountered in the study of Perot et al. [9]. A two-step calculation method is considered to avoid these problems. The computing efficiency of EGSnrc software [10] is far higher than MCNP5 code when used to calculate photon energy

*Supported by the National Natural Science Foundation of China (No.91226104) and National Special Magnetic Confinement Fusion Energy Research, China, under contract No.2015GB108001

1) *E-mail*: stingg@126.com

deposition. Therefore, we can simulate the prompt γ-ray spectrum using MCNP5, then simulate the ERS in liquid scintillator detector by EGSnrc code [11] using the calculated prompt γ-ray spectrum from MCNP5 as input.

In this paper, a two-step calculation method was used to simulate the energy deposition in a BC501A liquid scintillator detector for prompt γ-rays from a depleted uranium spherical shell bombarded by D–T neutrons. A coupled neutron photon transport calculation was carried out in the first step, while just photons in the second. In the first step, γ-rays below 7 MeV were simulated using MCNP5, while in the second step the ERS was simulated using EGSnrc. By comparing the calculated results with the experimental ones, the reliability of the two-step method in this work was checked.

## 2 Theory and methods
### 2.1 Physics of prompt γ-rays

Prompt γ-rays are emitted over a very short time interval ($10^{-14}$–$10^{-7}$ s) [12] through fission, inelastic reaction, and radiation capture for a depleted uranium spherical shell under D–T neutrons irradiation. The cross sections of the three processes for $^{238}$U are 0.58, $9.8\times10^{-4}$, and 1.1 barn, respectively. For the fission reaction, prompt γ-rays mainly come from the (n, f) reaction while the (n, fγ) reaction has little contribution. In the (n, f) reaction, prompt γ-rays are emitted from the excited compound nucleus or fission fragments when the excitation energy is inadequate to emit a neutron, namely fission prompt γ-rays. The energy of fission prompt γ-rays is 0.25–7 MeV, 95% of which are below 4.25 MeV. Binary fission of $^{238}$U is shown in Eq. (1), and there is a similar expression for the fission of $^{235}$U.

$$n + ^{238}U \rightarrow X + Y + \overline{\nu}n + \langle G \rangle \gamma + \langle E \rangle \tag{1}$$

In Eq. (1), X and Y represent fission fragments, $\overline{\nu}$ is the multiplicity of fission neutrons, and $\langle E \rangle$ and $\langle G \rangle$ are the average energy and multiplicity respectively of fission prompt γ-rays, which can be calculated by Eqs. (2) and (3) [13].

$$\langle E \rangle = -1.33(\pm 0.05) + 119.6(\pm 2.5) Z^{1/3}/A \tag{2}$$

$$\langle G \rangle = \frac{E_t(\overline{\nu}, Z, A)}{\langle E \rangle} \tag{3}$$

In Eq. (2), 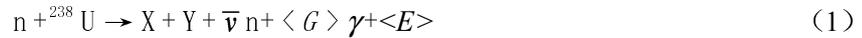 $\langle E \rangle$ is in MeV, $Z$ and $A$ are the atomic number and pre-fission mass of the fissionable nucleus, respectively. In Eq. (3), $E_t$ is the total prompt γ- ray energy.

For the inelastic reaction (n, n'γ), there are 88 energy levels for the excited $^{238}$U nucleus. Meanwhile, hundreds of γ-rays of 44.91 keV to 5.2 MeV are emitted and form a continuum γ-ray spectrum [14]. For the radiation capture reaction $^{238}$U (n, γ) $^{239}$U, there are 168 prompt γ-rays with various energies from 11.98 keV to 4.8 MeV. However, the contribution of γ-rays from radiation capture to the total is negligible as the capture cross section is about 0.1% of that for the fission reaction. Furthermore, fission neutrons of $^{238}$U and $^{235}$U could also induce the above-mentioned nuclear reactions and generate prompt γ-rays from collisions.

### 2.2 Simulation of prompt γ-rays

An anisotropic D–T neutron source was placed in the center of a depleted uranium spherical shell. A BC501A liquid scintillator detector was placed in the direction of the D$^+$ beam at a distance of 10.7 m from the neutron source. The schematic experimental layout is shown in Fig.1. Fig.1 (b)–(e) are enlarged diagrams of the depleted uranium spherical shell, collimator, liquid scintillator and target system including the titanium tritide target. The inner and outer diameters of the depleted uranium spherical shell are 50.8 cm and 57 cm, respectively. A thickness of 3.1 cm corresponds to 0.87 mean free path for 14 MeV neutrons. The depleted uranium was composed of $^{238}$U (99.579%), $^{235}$U (0.4154%), $^{234}$U (0.034%) and $^{236}$U (0.003); the density and total mass are 18.8 g/cm$^3$ and 526.6 kg, respectively.

Prompt γ-rays from scattered neutrons in the experiment hall were subtracted as background in the experiment, so it is not necessary to consider the experiment hall in the simulation model. The rectangular regions designated by dot-dash lines in Fig.1 (a) are the regions of the simulation models. Utilizing the coupled neutron-photon transport mode and tally type F2 in MCNP5 code, the prompt γ-ray energy spectrum and time spectrum were obtained on the surface "S1" (shown in Fig.1(c)) near the collimator. The energy cutoffs of neutron and photons were the default values in the simulation.

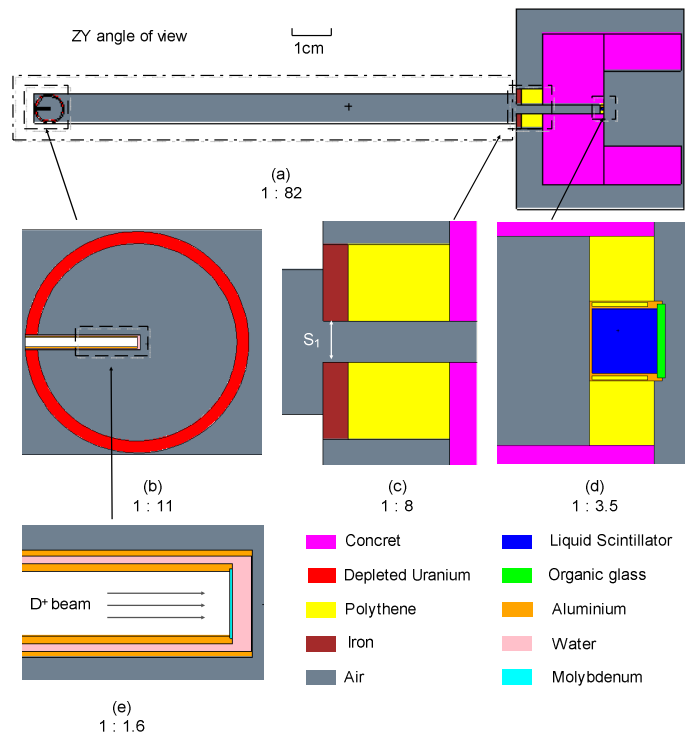

Fig.1. Experimental setup: (a) whole arrangement, (b) depleted uranium spherical shell, (c) collimator, (d) liquid scintillator detector, (e) target system.

## 2.3 Simulation of ERS

The EGSnrc system is a general-purpose package designed for MC simulation of the coupled transport of electrons and photons in arbitrary geometries for particles with energies above a few keV up to several hundreds of GeV. The system includes mainly source code and several user codes. Two codes, PEGS4 [11] and DOSRZnrc [11], were both used in this work. The following physical processes were considered: Compton scattering, photoelectric effect, pair production, bremsstrahlung production, positron annihilation, and Rayleigh scattering. The electron energy cutoff "ECUT" was 0.521 MeV (including the electron rest mass), and the photon energy cutoff "PCUT" was 1 keV.

PEGS4 was used to generate cross section data. Through modeling the passage of an electron or photon beam in a cylindrical geometry (shown in Fig.2), DOSRZnrc was used to simulate the energy deposition distributions induced by photons on surface "S1". The distance between the photon source (S1) and the front face of the liquid scintillator was 165 cm. The radius of the parallel photon beam was 3.15 cm, which is equal to the inner radius of the collimator. The spectrum of the incident photon source was the prompt γ-ray spectrum obtained by MCNP5, and the incident direction as normalized to Z-axis, which is perpendicular to the front face of the detector.

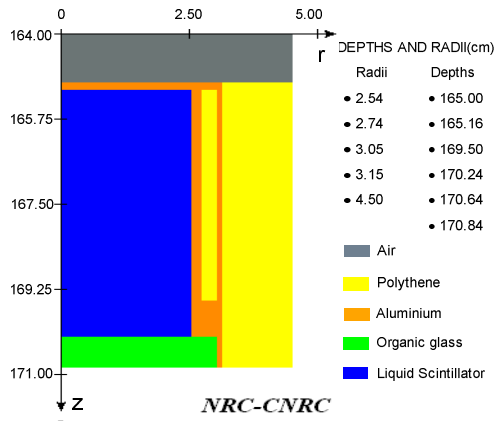

Fig.2. Simulation model of BC501A liquid scintillator detector and collimator.

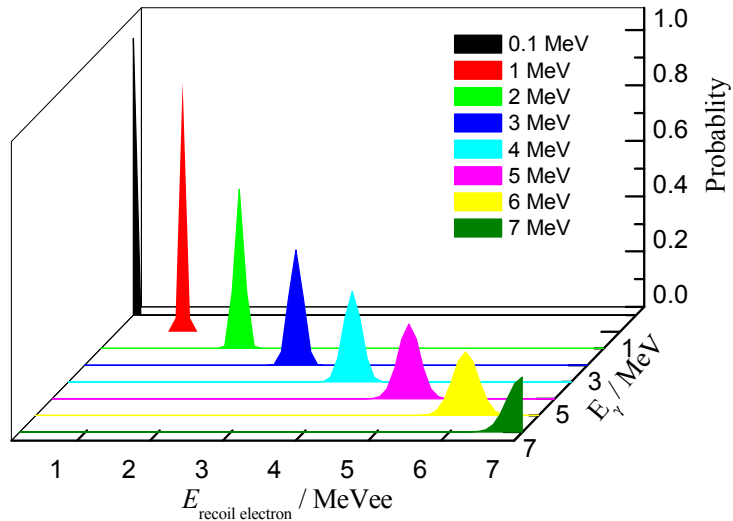

Fig.3. Energy resolution functions of BC501A liquid scintillator detector.

The energy resolution of the detector is not involved in the simulation. In order to compare with the experiment results, the simulated ERS needs to be broadened with a Gaussian function based on the measured energy resolution of the detector. A 70 × 70 Gaussian matrix was built based on Gaussian probability density function as shown in Eq. (4), the energy bins of which were 0.1 MeV steps from 0.1 MeV to 7 MeV. Some energy resolution functions of the BC501A detector are shown in Fig. 3. The broadened ERS could then be obtained from convolution of the simulated ERS with the Gaussian matrix.

$$\rho = \frac{1}{\sqrt{2\pi}\sigma} e^{-\frac{(E-E_0)^2}{2\sigma^2}} \quad (4)$$

In Eq. (4), $\sigma = \frac{1}{2\sqrt{2\ln 2}}(a + b\sqrt{E + cE^2})$, $E$ is in MeV. For the detector in this work, the values of a, b and c were -0.00102, 0.078014 and 0.608164, respectively.

## 3 Results from simulation
### 3.1 Energy spectrum and time spectrum of prompt γ-rays

In the MCNP5 simulation, a total count of $1.49 \times 10^9$ D–T neutrons was sampled. This pent 3000 minutes for a computer with 2.66 GHz CPU. The number and average energy of the photons in various cells are shown in Table 1. The energy and time spectra of prompt γ-rays on surface "S1" are shown in Fig. 4 and Fig. 5 respectively.

For the neutron cross sections used in MCNP5: $^{14}$N and $^{16}$O are from ENDF/VI.8 (actia), $^{238}$U and $^{235}$U are from the ENDF/BVI.5 Library (endf66c), and $^{234}$U and $^{236}$U are from the ENDF/BVI.0 Library (endf66c). For the photon and electron production cross sections, C, O and U are from the ENDF/BVI.8 Library (mcplib04 and el03). As the total enrichment of $^{234}$U and $^{236}$U is less than 0.05%, the contribution of the two nuclei can be neglected in practice.

Table 1. The number and average energy of photons in various cells.

| Cell | Number of photons | Average energy /MeV | Ratio % |
|---|---|---|---|
| Titanium film | $3.54 \times 10^5$ | 1.82 | $7.65 \times 10^{-3}$ |
| Molybdenum substrate | $5.60 \times 10^7$ | 1.68 | 1.21 |
| water | $2.69 \times 10^7$ | 4.84 | $5.82 \times 10^{-1}$ |
| Aluminum target chamber | $5.44 \times 10^7$ | 2.67 | 1.17 |
| Depleted uranium sphere | $4.49 \times 10^9$ | 0.65 | $9.70 \times 10^1$ |
| Air | $1.17 \times 10^5$ | 4.23 | $2.53 \times 10^{-4}$ |
| Total | $4.63 \times 10^9$ | — | 100 |

The results in Table 1 show that the average energy of photons is 0.65 MeV in the depleted uranium spherical cell. The simulated average energy for $^{238}$U is lower than the 0.97 MeV which was calculated by Eq. (2). The discrepancy in average energy indicates that the γ-rays from inelastic reaction may have been included in the simulated results, which have a lower average energy than those from the fission reaction. The average energy of photons in titanium film, molybdenum substrate, aluminum target tube and cooling water are shown in Table 1, and the probabilities of penetrating the 3.1 cm thick depleted uranium for these γ-rays are 0.120, 0.108, 0.164 and 0.196, respectively. Photons produced by air and neutrons can be neglected. Therefore, considering the average photon flux, we can conclude that the penetrating photons that are not from uranium occupy a ratio of about 3.6% of the total penetrating photons.

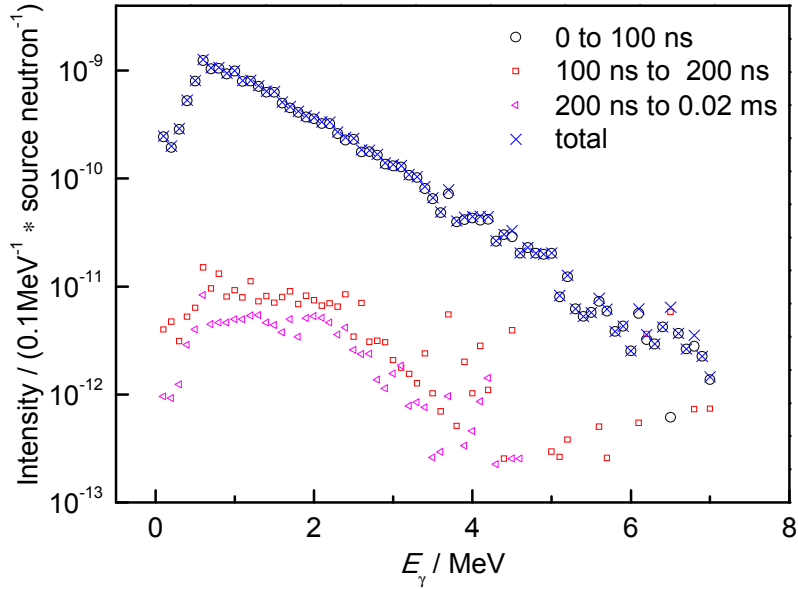

Fig. 4. Energy spectra of prompt γ-rays simulated by MCNP5

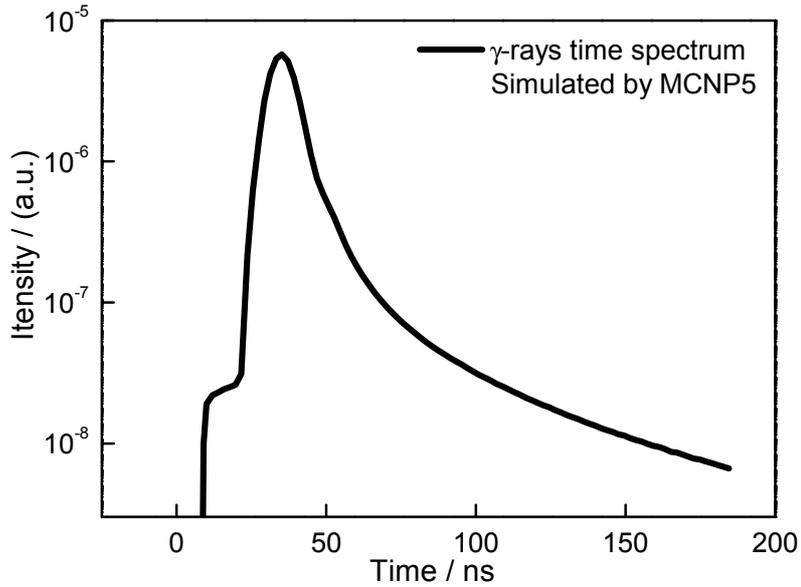

Fig. 5. Time spectrum of prompt γ-rays simulated by MCNP5.

The energy spectra of prompt γ-rays in different time intervals on the surface "S1" are shown in Fig. 4, with the emission process of most prompt γ-rays being finished within 100 nanoseconds. The statistical error in the prompt γ-rays is 2%–6% below 3 MeV, 6%–15% in 3–5 MeV, and it will reach 15% when the energy is above 5 MeV. The time spectrum of prompt γ-rays is shown in Fig. 5. 14 MeV D–T neutrons arriving at the inner surface of the depleted uranium spherical shell take about 5 ns, which corresponds to the starting point of the curve. The prompt γ-rays from the depleted uranium spherical shell arriving at the surface "S1" take about 29 ns, which corresponds to the second turn in the curve. Hence, it can be predicted that the γ-ray intensity should reach a maximum at 34 ns, and then decrease sharply. The γ-ray flux should decrease two orders of magnitude by 100 ns. Therefore, a period of 100 ns for the electronic circuit is appropriate in this experiment.

**3.2 Electron recoil spectrum in detector**

Utilizing the prompt γ-ray spectrum from the MCNP calculation as input, the simulated ERS by EGSnrc and the broadened ERS for the BC501A detector are shown in Fig. 6. The total efficiency of the continuum γ-rays is 0.1914. Results indicate that the shapes of the two ERS spectra are consistent with each other over the entire energy region, while the shape of the broadened ERS is much smoother than that of the original one beyond 5 MeV.

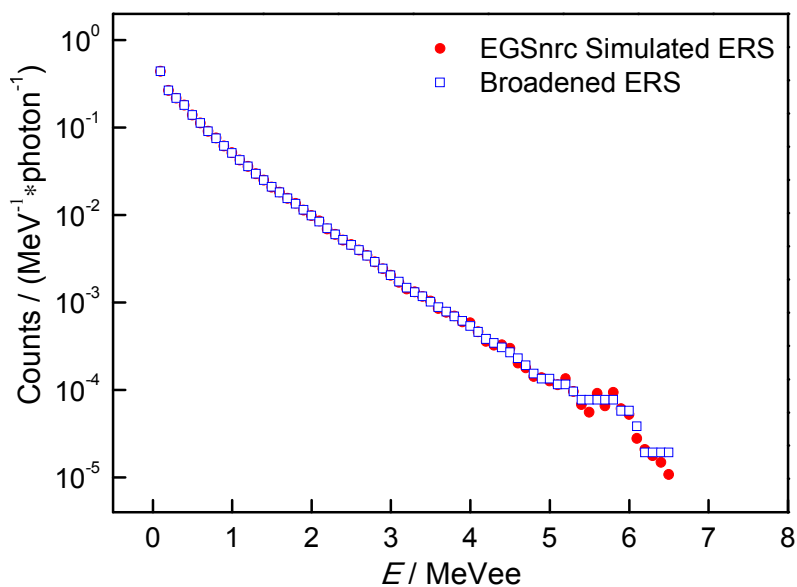

Fig.6. Simulated and broadened ERS.

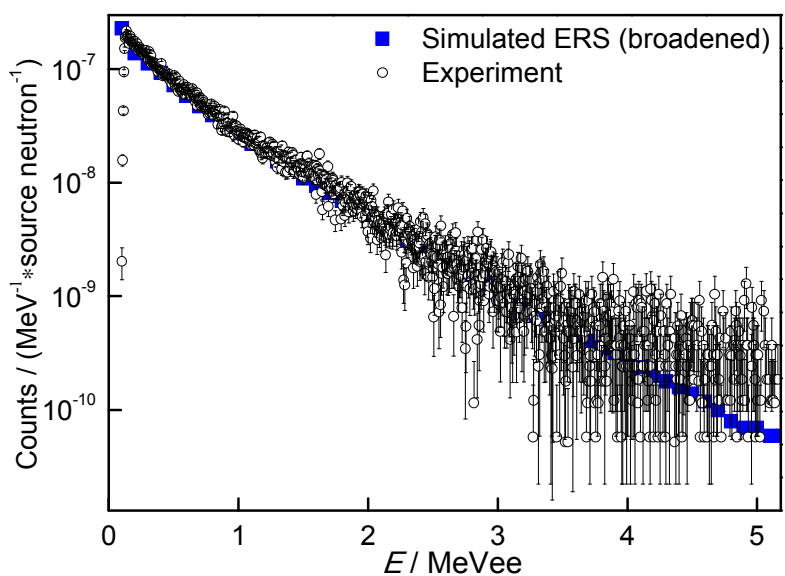

Fig.7. Comparison of simulated ERS to experimental results.

**4 Experimental validation**

Validation of the ERS and energy spectrum of prompt γ-rays by Monte Carlo simulation was accomplished experimentally. Based on a pulsed D–T neutron source from the PD–300 neutron generator at the Institute of Nuclear Physics and Chemistry (INPC), Mianyang, the ERS for prompt γ-rays from a depleted uranium spherical shell were obtained by the ToF method. The detector in simulation and experiment was exactly the same. The ERS was unfolded by an iterative method for γ-rays below 4 MeVee, and then the prompt γ-ray spectrum was obtained.

The ERS from simulation and experiment are shown in Fig. 7. The two spectra agree with each other in the range 0.15–4 MeVee, while there are insufficient statistics above 4 MeVee. The unfolded prompt γ-ray spectrum below 4 MeV is shown in Fig. 8, and the experimental results agree well with the simulation in the range 0.4–3 MeV.

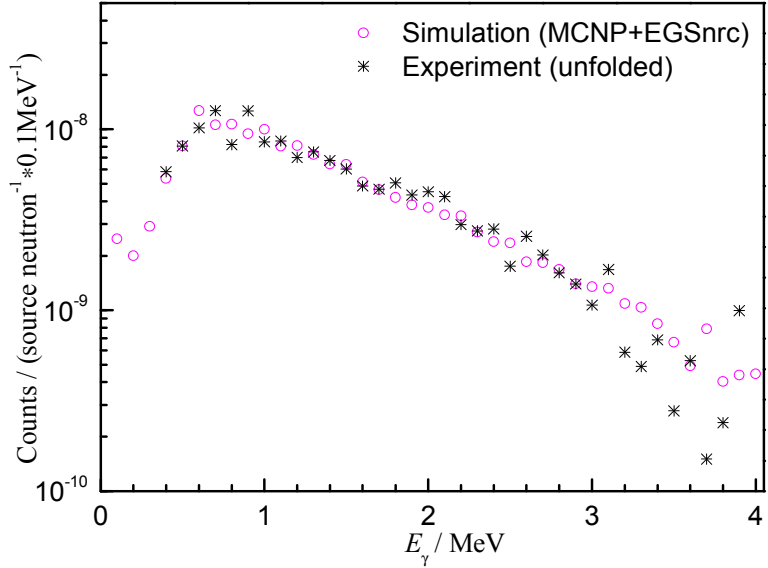

Fig.8 Comparison of simulated prompt γ-ray spectra and experimental results.

## 5 Conclusion

Prompt γ-ray spectra from a depleted uranium spherical shell irradiated by D–T neutrons and the corresponding ERS in a BC501A liquid scintillator detector have been quantitatively specified with the Monte Carlo codes MCNP5 and EGSnrc. The ERS and γ-ray spectrum from simulations were compared with the experimental results. The general agreement between experiment and simulation is quite satisfactory in the energy region of 0.4–3 MeV for the prompt γ-ray spectrum, and below 4 MeVee for ERS. The two-step calculation method provided in this work is efficient and could avoid the unreliability of pulse height tally in energy deposition of γ-rays under neutron irradiation. It can be confirmed that the two-step simulation method is reliable and accurate, and can be applied to the simulation of prompt γ-ray energy deposition in neutron γ-ray mixed fields.

**Acknowledgements**

We would like to thank Mou Yunfeng and Wang Xinhua for their helpful valuable discussions. We are indebted to the reviewers of this paper for their constructive remarks.